# THE PHASE DIAGRAM OF HIGH-$T_C$ SUPERCONDUCTORS IN THE PRESENCE OF DYNAMIC STRIPES

Annette Bussmann-Holder, Alan R. Bishop[1], Helmut Büttner[2],
Takeshi Egami[3], Roman Micnas[4] and K. Alex Müller[5]

Max-Planck-Institut für Festkörperforschung, Heisenbergstr.1, D-70569 Stuttgart, Germany
[1] Theoretical Division, Los Alamos National Laboratory, Los Alamos, NM 87545, USA
[2] Lehrstuhl für Theoretische Physik I, Universität Bayreuth, D-95440 Bayreuth, Germany
[3] Department of Materials Science, University of Pennsylvania, LRSM-3231 Walnut Street Philadelphia, PA 19104, USA
[4] Institute of Physics, A. Mickiewicz University, 85 Umultowska St., 61-614 Poznan, Poland
[5] Physik Institut Universität Zürich, Winterthurerstr.190, CH-8057 Zürich, Switzerland

The phase diagram of superconducting copper oxides is calculated as a function of doping based on a theory of dynamic stripe induced superconductivity. The two major conclusions from the theory and the numerical analysis are that T* (the pseudogap onset) and $T_c$ (superconductivity onset) are correlated through the pseudogap, which induces a gap in the single particle energies that persists into the superconducting state. By decreasing the doping the pseudogap Δ* increases and T* increases, but when Δ* exceeds a certain critical value the superconducting transition temperature $T_c$ is suppressed. A mixed s- and d-wave pairing symmetry is also examined as function of doping.

The phase diagram of high temperature superconducting copper oxides [1] exhibits, as a function of hole doping, an unusual richness. It shows antiferromagnetism in the underdoped regime, a "strange" metal state, dominated by inhomogeneous charge distribution (e. g. stripes), and a superconducting state with unexpectedly high transition temperatures. At doping levels beyond the superconducting phase a metallic behaviour is observed.

The understanding of this diagram is at present very incomplete and few concepts exist to address the full range of experimentally observed properties. In particular, the antiferromagnetic properties have challenged theoreticians to explain the pairing mechanism and the phase diagram in terms of purely electronic models based on electron correlation and/or spin fluctuation mechanisms [2]. In these approaches the onset temperature T* of the so-called pseudogap opening is usually identified as a spin gap temperature, even though various recent experiments have revealed a giant oxygen and copper isotope effect on T* [3, 4, 5]. These findings point to strong lattice effects being important in the pseudogap formation. Correlated with the above isotope effects are experimental results obtained from EXAFS [6], inelastic neutron scattering [7], NMR [8] and EPR [9], which all show that strong anomalies in the lattice take place at T*. These results suggest that in order to model high-temperature superconductors and understand their phase diagram, a purely electronic mechanism is insufficient. Rather the interaction between spin, charge and phonons (lattice) has to be properly included. Another crucial issue, which is also debated experimentally, is the presence or absence of a correlation between T* and $T_c$ (the superconducting transition temperature). While various approaches exist which invoke the striped phase as destroying superconducitivity [10], others assume a positive effect of T* on $T_c$ [11]. Also it is strongly debated how to define the phase separating T* and $T_c$. While tunneling data [12] support a coincidence between the optimum $T_c$ and T*, other tunneling experiments [13], together with those from [3, 4], find that T* is larger than $T_c$ at the optimum doping. This controversial issue has the important consequence that some theories support the possibility of a quantum critical point [14] (QCP) inside the superconducting phase, while the latter data give no evidence for this scenario but rather only a 2D/3D crossover of the quantum critical XY model [15],

In the present approach we adopt explicitly in our Hamiltonian effects stemming from out-of-plane orbitals, e. g., Cu $d_{3z^2-r^2}$ and oxgen $p_z$ states, in addition to in-plane Cu $d_{x^2-y^2}$ and oxygen $p_{x,y}$ bands [16, 17]. In the undoped antiferromagnetic parent compounds, symmetry considerations do not admit for direct hopping processes between the in-plane and out-of-plane components. With doping, strain inducing plane buckling / octahedra tilting sets in [18], which dynamically lowers the symmetry and produces hybridization between e.g. $d_{x^2-y^2}$ and $p_z$ and $d_{3z^2-r^2}$ and $p_{x,y}$ electronic states. Importantly the buckling / tilting induces strong electron-phonon interaction processes, which lead to the appearance of inhomogeneously modulated phases, e.g. stripes, in spin, charge and lattice [19]. This doping-induced buckling / tilting may also be a natural mechanism for a crossover from 2 to 3 dimensions with doping [15]. An effective two-band Hamiltonian can be obtained by introducing from the beginning in-plane and out-of-plane elements represented by strongly p-d hybridized bands,. This corresponds to a two component scenario [20], wherein one component is related to an incipient spin channel, while the other is identified as an incipient charge channel. As outlined above, the coupling between these components is due to the lattice distortion (buckling).

The lattice renormalized two-band Hamiltonian is given by:

$$H = \sum_{i,\sigma} E_{xy,i} c^+_{xy,i,\sigma} c_{xy,i,\sigma} + \sum_{j,\sigma} E_{z,j} c^+_{z,j,\sigma} c_{z,j,\sigma} + \sum_{i,j,\sigma,\sigma'} T_{xy,z} [c^+_{xy,i,\sigma} c_{z,j,\sigma'} + h.c.] + \sum_{i,j} \tilde{T}_{xy} n_{xy,i\uparrow} n_{xy,j\downarrow}$$
$$+ \sum_{i,j,\sigma,\sigma'} \tilde{V}_C n_{xy,i,\sigma} n_{z,j,\sigma'} + \sum_{i,j,\sigma,\sigma'} V_{pd} n_{xy,i,\sigma} n_{xy,j,\sigma'} ,$$

(1)

where $c^+_{xy,i,\sigma} c_{xy,i,\sigma} = n_{xy,\sigma,i}$ and $c^+_{z,j,\sigma} c_{z,j,\sigma} = n_{z,j,\sigma}$ are the site i,j-dependent plane and c-axis (z) electron densities with single particle energy $E$ and spin index $\sigma$. $T_{xy,z}$ is the hopping integral between plane and c-axis orbitals, and $\tilde{T}_{xy}$ is the in-plane phonon renormalized spin singlet extended Hubbard term from which a d-wave symmetry of a superconducting order parameter would result [21]. $\tilde{V}_C$ as well as $V_{pd}$ are density-density interaction terms referring, respectively, to plane / c-axis and in-plane elements. The phonon contributions have already been incorporated at an adiabatic level in equation (1), so that all energies given are renormalized quantities [16]. Since the phonon renormalization has been discussed in detail in [16], only their effect on the single particle energies will be repeated here, viz:

$$E_{xy,i} = [\varepsilon_{xy,i} - (g_i^{(xy)} Q_l^{(xy)} - \tilde{g}_{i,m}^{(xy,z)} Q_m^{(z)} <n_{z,m}>)\ ] =$$
$$[\varepsilon_{xy,i} - \{\Delta_i^{xy} + f(xy,z)\}\ ]$$
$$E_{z,j} = [\varepsilon_{z,j} - (g_j^{(z)} Q_m^{(z)} - \tilde{g}_{j,l}^{(xy,z)} Q_l^{(xy)} <n_{xy,l}>)\ ] =$$
$$[\varepsilon_{z,j} - \{\Delta_j^z + f(z,xy)\}\ ]\ .$$

(2)

Here $\varepsilon$ are the unrenormalized site (i,j) dependent band energies of the xy and z related hybridized bands and the $g$'s are in-plane (xy) and out-of-plane (z) electron-phonon couplings with the corresponding site ($l, m$) dependent averaged phonon displacements $Q$. The electron-phonon couplings $\tilde{g}$ refer to in-plane / out-of-plane couplings with strong charge transfer character and band mixing. The phonon mode energies which are considered here, are, for the in-plane elements, the $Q_2$-type LO mode which shows anomalous dependences on temperature and composition and has been shown to interact strongly with charge [22, 23, 24], while the out-of-plane mode is the low energy polar mode, which also shows many anomalous properties [25]. It is clear from equation (2) that two instabilities may occur due to the spin-charge-lattice coupling, since gaps proportional to $\Delta$ are induced in the single particle energies: a charge density wave instability can be induced in the charge channel, while a spin density wave instability can occur in the spin channel. Both instabilities are partially suppressed due to the third terms in the renormalized band energies, which stems from the coupling of in-plane and out-of-plane orbitals due to buckling / tilting, corresponding to a strongly anharmonic local mode. This phonon mediated interband interaction term provides dispersion to the quasi – localized spin channel and flattens dispersion in the metallic type charge channel. In addition it mediates the dynamical character of charge / spin ordering related to dynamical stripe formation due to the local character of this buckling / tilting mode. Note that the incipient spin and charge density wave instabilities both have mixed in-plane and c-axis character due to the buckling- induced coupling.

In view of the observation of a huge isotope effect on T*, [3, 4, 5] we relate T* to the opening of an incipient dynamical charge gap, given by $\Delta^z$. In [19] we have shown that this isotope effect is indeed captured within the above model. Here we calculate the corresponding transition temperature T* within the meanfield framework of [26]. We effectively include doping in this approach through the variation of the Fermi energy, T* and the corresponding $\Delta_z = \Delta*$ as

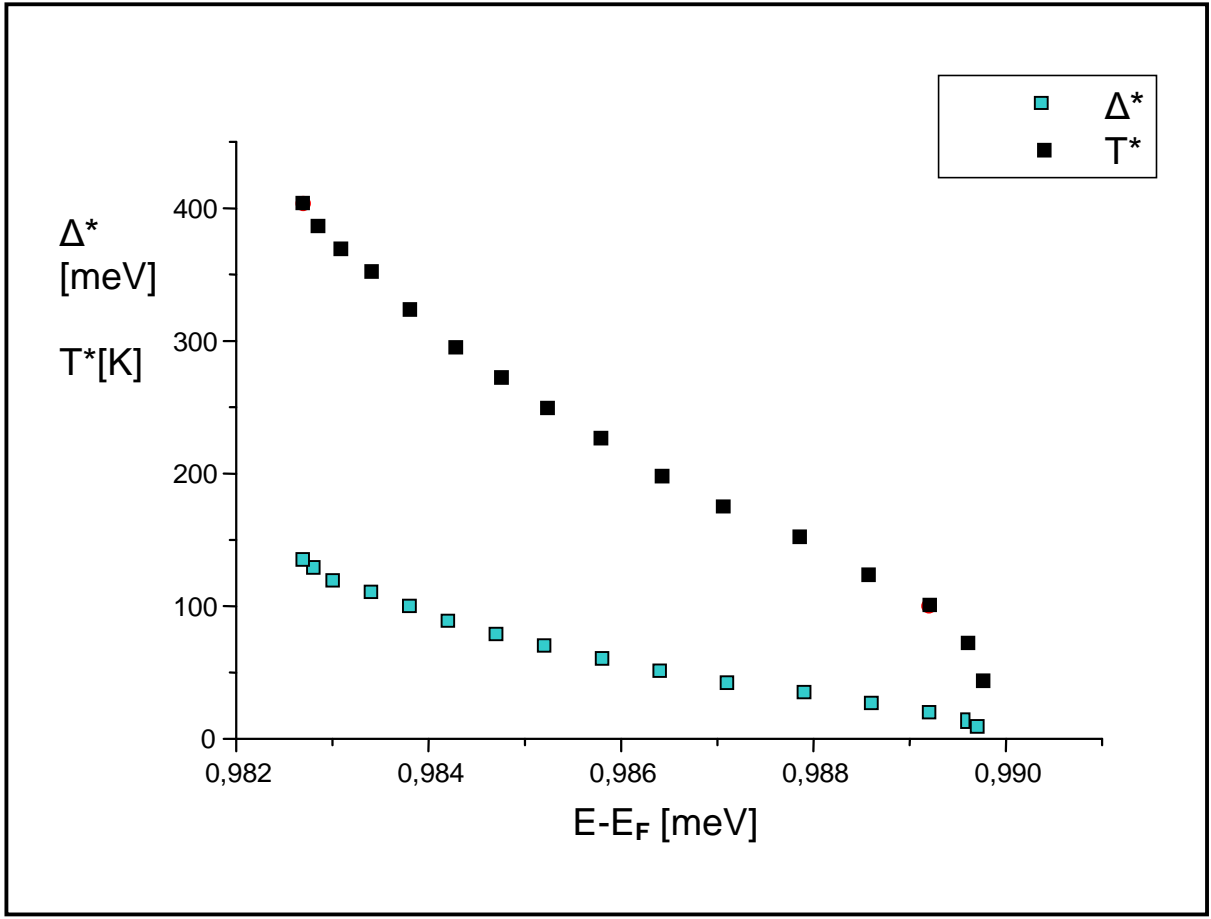

Figure 1: The pseudogap $\Delta*$ and its onset temperature T* as functions of doping. The parameters used are, within the formalism introduced in Ref. 16, $g_j^{(z)} Q_m^{(z)} = 18.5 meV$ at T=400K. The temperature dependences of T* and $\Delta*$ are evaluated selfconsistently; the spatial (stripe) modulations are not included here.

functions of doping are shown in fig. 1. Here the absolute magnitude of T* depends on the corresponding phonon energy which renormalizes, hardening by approximately 2% with increased doping. The parameters used in this calculation are given in figure caption 1. Simultaneously, the zero temperature gap has been deduced using the scheme of ref. [26]: As we have shown recently [16], the (dynamical) gaps in the single particle site energies have the important effect of providing a "glue" between the two components, with the consequence of enhancing $T_c$ dramatically to the experimentally observed values even if both components are – when uncoupled – not superconducting. This important result can be understood in the following way (Figure 2):

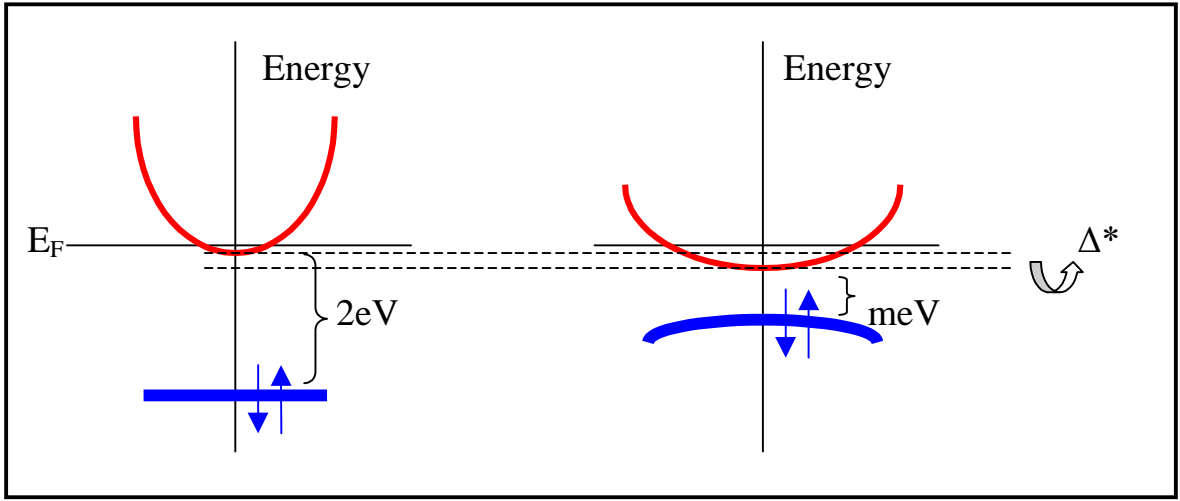

Figure 2: Schematic energy level structure of the charge and spin related components. Red line refers to the charge channel, blue line to the spin channel. Left panel shows the energy level structure in the absence of the pseudogap $\Delta^*$, while the right panel shows the effect of $\Delta^*$ on the level positions and dispersions.

The spin related channel is characterized by the formation of a singlet state [21, 27] which is well below the Fermi energy and highly localized due to nearby triplet states and the squeezing effect arising from the coupling to the $Q_2$-type phonon mode. The effect of doping and the "spin" gap is to shift this state towards the Fermi level and to provide mobilty through its dispersion. The charge related channel has Fermi liquid like properties, is highly mobile and located very near to the Fermi energy. Here the charge gap shifts the states towards the other singlet related band and reduces dispersion and hence the mobility. Since both effects combine to drive the charge and spin channels together, interband processes are easily facilitated. It is just this interband coupling driven by buckling / tilting which has, in addition to the single particle gaps, an important enhancement effect on $T_c$. From the knowledge of both the dependence of T* and $\Delta_z = \Delta^*$ on doping, and the dependence of $T_c$ on $\Delta^*$, the predicted phase diagram for the cuprates can be deduced (Fig. 3). It is clear from fig. 3, that the maximum $T_c$ at optimum doping is smaller than the corresponding T*. Both temperature scales vanish simultaneously in the overdoped regime. It is interesting to note that the absolute increase in T* with isotopic substitution decreases with increasing doping. There is a complementary absolute decrease in $T_c$ which also decreases with increasing doping for complete isotopic substitution [29]. The site- selective dependence of these isotopic shifts [30] is mainly determined by associated densities of states, and dominated by the planar contributions [31].

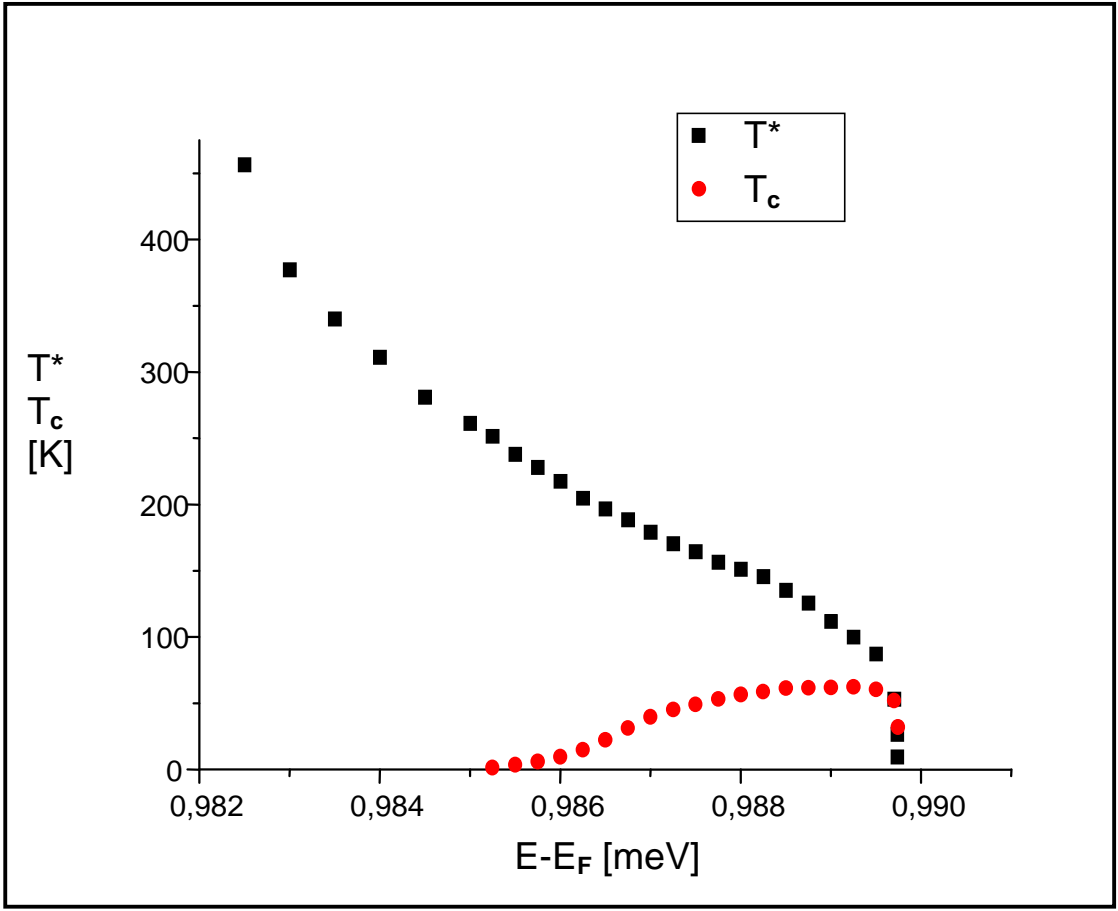

Figure 3: The calculated phase diagram of HTSC: black squares represent the onset temperature T* of the pseudogap formation, red circles give the superconducting transition temperature $T_c$. Using the formalism of Ref. 16, $T_c$ is calculated within the framework of the standard two-band model approach of Suhl et al. [28] with dimensionless parameters $V_{im}=V_{jn}=-0.01$ and $V_{ijmn}=0.4$. (Note that the parameters used include the appropriate densities of states.)

Importantly, the present calculation establishes a direct connection between $T_c$ and T* and $\Delta_{sc}$ and $\Delta^*$, and yields a mixed pairing symmetry. In figure 4 we show the calculated dependences of $2\Delta_{sc,max}(d)/kT_c$ and $2\Delta_{sc,max}(s)/kT_c$ on $T^*/T_c$, where we relate the d-wave component to in-plane pairing and the s-wave component to the out-of-plane pairing, respectively. These dependences are approximately linear. As a function of increased doping the relative s-wave contribution increases.

In conclusion, we have analysed a two-component model of high-temperature superconducting cuprates in which the two components are coupled by buckling / tilting of the $CuO_2$ planes. The two components considered are in-plane and out-of-plane structural elements. The coupling between these elements occurs through incipient spin- and charge-densitiy-wave instabilities and leads to high-temperature superconductivity with mixed s- and d-wave pairing symmetry. The predicted pseudogap temperature T* is always greater than $T_c$ so that there is no real quantum critical point within the superconducting region Extensions of the present approach are underway to explicitly include effects of the spatial charge/spin inhomogeneity supported by equ.1 [19].

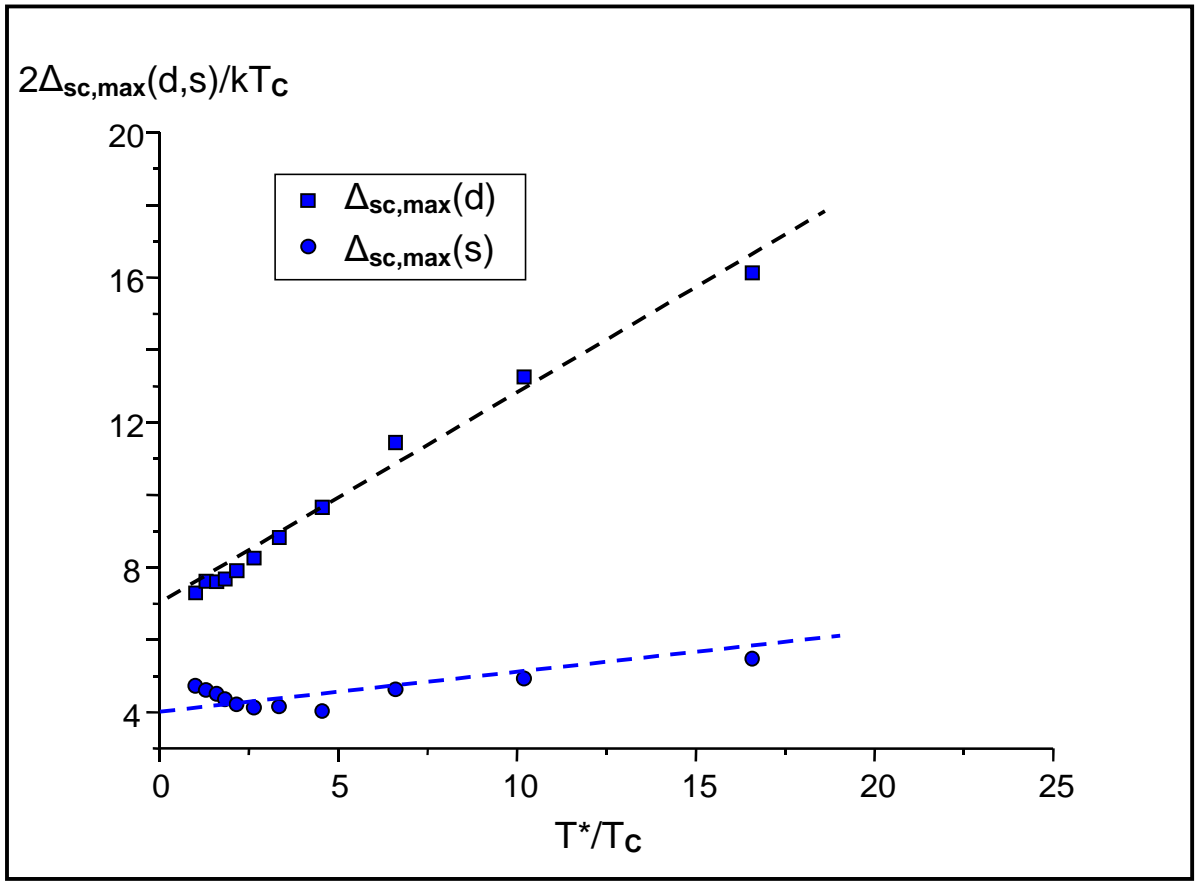

Fig. 4: $2\Delta_{sc,max}(s,d)/kT_c$ as function of $T^*/T_c$ for doping levels >0.986 (E-E$_F$). The lines are guides to the eye. Note that the s-wave component gains weight with increasing doping.